\documentclass[aps,reprint,amsmath,amssymb,pra,twocolumn]{revtex4-2}
\usepackage{graphicx}
\usepackage{dcolumn}
\usepackage{bm}
\usepackage{float}
\usepackage{cancel}
\usepackage{xcolor}
\usepackage{lipsum}
\usepackage{lineno}

\newcommand{\p}{\partial}
\newcommand{\Omeg}{\textsl{g}}

\newcommand{\vta}{\vartheta}
\newcommand{\om}{\omega}
\newcommand{\vg}{\textsl{g}}
\newcommand{\vD}{\varDelta}
\newcommand{\nn}{\nonumber}
\newcommand{\ta}{\theta}
\newcommand{\sq}[1]{\sqrt{\smash[b]{#1}}}
\newcommand{\al}{\alpha}
\newcommand{\wt}{\widetilde}
\newcommand{\wh}{\widehat}

\newcommand{\cR}{{\cal R}}
\newcommand{\cH}{{\cal H}}
\newcommand{\cF}{{\cal F}}

\newcommand{\cW}{{\cal W}}

\newcommand{\textblue}[1]{\textcolor{blue}{#1}}
\newcommand{\be}{\begin{equation}}                                       
\newcommand{\ee}{\end{equation}}
\newcommand{\ba}{\begin{eqnarray}}
\newcommand{\ea}{\end{eqnarray}}
\newcommand{\bref}[1]{(\ref{#1})}

\newcommand{\lab}[1]{\label{#1}}

\newcommand{\bsub}{\begin{linenomath}\begin{subequations}}                      
\newcommand{\esub}{\end{subequations}\end{linenomath}}

\begin{document}
	
	\preprint{APS/123-QED}
	
\title{Finesse and four-wave mixing in microresonators}
\author{D.N. Puzyrev}
\author{D.V. Skryabin}
\email{d.v.skryabin@bath.ac.uk}
\affiliation{Department of Physics, University of Bath, Bath BA2 7AY, UK}%

\begin{abstract}
We elaborate a comprehensive theory of the sharp  variations of the four-wave-mixing (FWM)
threshold happening with the tuning of the pump frequency along the dispersive tails of the nonlinear resonance in the driven Kerr microresonators with high finesses and high finesse dispersions. Our theory leads to the explicit estimates for the difference in the pump powers required for the excitation of one sideband pair with  momenta $\pm\mu$, and for the simultaneous excitation of the  two neighbouring pairs with momenta $\pm\mu$, $\pm(\mu+1)$. This power difference also measures the depth of the Arnold tongues in the pump-frequency and pump-power parameter space, associated with the aforementioned power variations.
A set of select pump frequencies and powers, where the instabilities of the two sideband pairs come first, forms the threshold of complexity, which is followed by a sequence of conditions specifying critical powers for the simultaneous instabilities of three, four and so on pairs. 
 We formally link finesse dispersion and the density of states notion borrowed from the condensed matter context, and report a value of the finesse dispersion and a critical mode number signalling the transition from the low- to the high-contrast tongue structure.
 We also demonstrate that the large finesse dispersion makes possible a surprising for the multimode resonators regime of the bistability without FWM. 
\end{abstract}
	\maketitle
	

\section{Introduction}
Maxwell and Schroedinger equations are the fundamental laws of nature
governing  interaction of photons and electrons.
Optically pumped resonators boost the number of incoming photons interacting with the intra-resonator matter  by a factor determined by the inter-mode frequency separation divided by the linewidth, i.e., finesse. Thereby, the high-finesse resonators 
are the devices enabling modern research into strong light-matter interaction~\cite{rev5} and frequency conversion~\cite{rev4}. In this work, we present a comprehensive analytical theory for the four-wave mixing (FWM) threshold in the high-finesse Kerr resonators. Our research is underpinned by the relentless push towards achieving ever-high quality factors in the bulk and chip-integrated microresonators, 
see, e.g.,~\cite{hq1,hq2,hq3,hq4,hq5}.

Since the very early days of theoretical research into nonlinear physics of optical resonators, it has been acknowledged that the Maxwell and Schroedinger equations can be reduced to either a system of coupled ordinary differential equations (ode's) for the time evolution of mode amplitudes~\cite{lamb} or  a system of partial differential equations (pde's) for the envelope functions~\cite{risk2}. 
Two approaches are in fact equivalent providing a pde is supplemented by the boundary conditions that admit the mode profiles used to derive the ode's \cite{risk2,llprl0}. A high-quality ring microresonator~\cite{rev4,rev1} is an example   where the aforementioned equivalence is particularly vivid, since the  azimuthally rotating modes, $e^{i\mu\vta-i\om_\mu t}$, constitute a quasi-exact basis for nonlinear Maxwell equations and the exact one for their reduction to the Lugiato-Lefever equation (LLE) conditioned by the $2\pi-$periodicity in $\vta$~\cite{chembo,herr,optcom,mil,chembo1}, see, e.g., Refs.~\cite{revv2} for general and Ref.~\cite{cont} for the derivation focused overviews.

Thus, either the pde or ode formulations allow investigating the impact of the 
finesse on FWM and other types of frequency conversion in microresonators. Recently, we have contributed to the microresonator theory by demonstrating that the threshold condition for the degenerate FWM  in the high-finesse devices  breaks the pump laser parameter space into a sequence of narrow in frequency and broad in power  Arnold tongues~\cite{arnold}. Large finesse $\cF$ is normally accompanied by the relatively large finesse dispersion, $\cF_d$, which is a key parameter behind the tongue formation. Instability tongues become a dominant feature in resonators with  $\cF_d$ approaching unity and disappear if $\cF_d$ is relatively small. 
The latter case recovers the multimode threshold shapes reported in Refs.~\cite{chembo0,chembo1}. Ref.~\cite{chembo0} followed a series of pioneering  
experiments on FWM in  microresonators \cite{f1,f2,f3} and the three-mode, i.e., pump-signal-idler, theory of the microresonator FWM \cite{f4}.

Ref.~\cite{arnold} has addressed few interesting problems with the FWM emerging in the ultrahigh-$\cF$ devices. First, is that looking at the superposition of the individual thresholds for generation of the $\pm\mu$ sideband pairs, one can observe dramatic reshaping of the net threshold when the finesse dispersion, $\cF_d$, is increased.  Second, is  that the 
intersections of the instabilities domains corresponding to $\pm\mu$ and $\pm(\mu+1)$ modes
touch the stability intervals of the single-mode state and therefore constitute a threshold condition - {\em threshold of complexity}. These effects are happening 
along the dispersive tails of the nonlinear resonance for positive detunings if  dispersion is normal, $\cF_d<0$, and for the negative detunings if  dispersion is anomalous, $\cF_d>0$.
Elegant mathematics  and important physics behind deriving the power and detuning values corresponding to the threshold of complexity could not be given the attention they deserve in the short format of Ref.~\cite{arnold} and their detailed description is presented to the readers below. We also cover here a range of connected problems not directly addressed by Ref. \cite{arnold}. 
This work is organised in several sections and multiple subsections:  

Section~II introduces a model (II.A), notations
and parameters, including the  residual finesse,~$\al_\mu=(\mu+\tfrac{1}{2})\cF_d$, which is the key parameter in our study~(II.B). (II.C)~introduces transparent power scaling
allowing an immediate link to physical quantities of interest. (II.D)~covers in details 
the classic results~\cite{llprl} on the cw-state stability using relevant for us notations and terminology. It sets the scene for revealing the role of the interplay between the different $|\mu|$ in the cw-stability analysis, which constitutes the main content of the following sections. 

Section~III provides comprehensive description of  the visualised numerical data on the  interplay between the instabilities for different $|\mu|$, and of the threshold 
reshaping for the residual spectrum changing from the quasi-continuous to sparse. 
It takes an advantage of using the side-band coupling constant, $\vg$, for these purposes, and discusses details of nonlinear mapping between $\vg$, which is proportional to the intra-resonator power, 
and the pump laser power.

Section~IV is central. It solves conditions on $\vg$, along the threshold of complexity (IV.A, IV.B). It then works out critical values for $\mu=\wh\mu$, $\al_\mu=\al_{\wh\mu}$, and for the laser detuning that can be used to characterise transition to the Arnold-tongue regime (IV.C). IV.D  derives the $|\cF_d|> 2/\sqrt{3}$ condition for the Arnold tongues to dominate for all detunings, and demonstrates  an unusual for the multimode resonators regime of the bistability without FWM.

Section~V presents comprehensive explicit analytical results for the sibeband coupling constant and laser power along the threshold of complexity. It provides an overarching
diagram characterising occurrences of the different FWM instabilities 
in the high-finesse resonators.

Section VI provides further contextual links.  Section VII summarizes the results.

\section{Background}
\subsection{Microresonator LLE (mLLE)}
The mLLE model  is~\cite{revv2,cont} 
\bsub
\lab{ll}
\begin{align}
& i\p_t\psi=\delta_0\psi-\tfrac{1}{2}D_2\p_\ta^2\psi-i\tfrac{1}{2}\kappa
\left(\psi-\cH \right) 
-\gamma |\psi|^2\psi,
\lab{lla}\\
& \ta=\vartheta-D_1t,\lab{llb}
\end{align}
\esub
where 
$\delta_0=\om_0-\om_p$ is detuning of the pump laser frequency, $\om_p$,
from the $\om_0$ resonance.  $\vta$ is the  angular coordinate in the laboratory reference frame.
$D_1/2\pi>0$ is the non-dispersive part of the resonator repetition rate, which equals 
the nondispersive part of the FSR. $D_2$ is the 2nd order dispersion. $\kappa$ is the linewidth. 
$\cH^2$ is the intra-resonator pump power.
$\gamma$ is the nonlinear parameter~\cite{cont},
\be
\gamma= \frac{\om_0n_2}{2Sn_0},
\ee
where $n_2$ is Kerr coefficient, $n_0$ is refractive index, and $S$ is the transverse mode 
area.

Transformation to the rotating reference frame, $\vartheta\to\ta$, 
and counting  resonances from $\om_0$
replaces the  bare resonator spectrum
\begin{align}
&\om_\mu=\om_0+D_1\mu+\tfrac{1}{2}D_2\mu^2,
\lab{om}\\
\nn &\mu=-\tfrac{1}{2}N+1,\dots,0,\dots,\tfrac{1}{2}N,
\end{align}
with the residual, i.e., linear mLLE, spectrum
\be
\vD_\mu=\delta_0+\tfrac{1}{2}\mu^2D_2.\lab{resa}
\ee

Let us make a note, that all the theory is developed here in physical units. In particular, $\delta_0$, $\kappa$, $D_2$ have units of Hz. $|\psi|^2$, $\cH^2$ are measured in Watts, and $\gamma$ is in Hz/Watt.

\subsection{Finesse, residual finesse and density of states}
Resonator finesse, $\cF_\mu$, is a relative to the linewidth 
and the mode-number specific measure of the resonance separation, 
\bsub
\lab{fn}
\begin{align}
&\frac{\om_{\mu+1}-\om_\mu}{\kappa}= \cF_\mu=\frac{D_1}{\kappa}+\frac{D_2(\mu+\tfrac{1}{2})}{\kappa},~ \mu\ge 0,\lab{fna}\\
&\frac{\om_{\mu}-\om_{\mu-1}}{\kappa}= \cF_\mu=\frac{D_1}{\kappa}
+\frac{D_2(\mu-\tfrac{1}{2})}{\kappa},~ \mu\le 0.\lab{fnb}
\end{align}
\esub
Here,
\be
\cF=\frac{D_1}{\kappa},~~\text{and}~~\cF_d=\frac{D_2}{\kappa}
\lab{ff}
\ee
are the dispersion free finesse and finesse dispersion, respectively.

Following Ref.~\cite{arnold}, we define a rule to calculate frequencies in the residual spectrum,
\bsub
\lab{rf}
\begin{align}
&\frac{\vD_{\mu+1}-\vD_\mu}{\kappa}=\alpha_\mu,~
\al_\mu=\cF_d(\mu+\tfrac{1}{2}),~ \mu\ge 0,\lab{rfa}\\
&\frac{\vD_{\mu}-\vD_{\mu-1}}{\kappa}= \alpha_\mu,~\al_\mu=\cF_d(\mu-\tfrac{1}{2}),~ \mu\le 0.\lab{rfb}
\end{align}
\esub
Here, $\al_\mu$ is the residual finesse defining
the mode-number specific separation of the resonances in the residual spectrum \cite{arnold}.
Residual spectrum is non-equidistant and the resonance separation is increasing 
with $|\mu|$. $\cF_d$ is the prime and only parameter controlling this nonequidistance. 
$|\alpha_\mu|\ll 1$ corresponds to the quasi-continuous residual spectrum and 
$|\alpha_\mu|\gg 1$ to the sparse one.
For microresonators, $\cF$ is typically the order of $10^3$ to $10^6$, while $|\cF_d|$
is the order of $10^{-3}$ to $10$, so that $\cF\gg|\cF_d|$.

Definition of the residual finesse directly connects to the  density of optical states (DoS)  concept~\cite{dos1,dos2}. Taking the ${\boldsymbol{\mathcal L}}_\mu=2|\mu|$ length span in the momentum space, we apply the text-book definition of the density of states~\cite{book2}, ${\boldsymbol{\mathcal D}}_\mu$, to the $\kappa$-normalised residual spectrum, 
\be
{\boldsymbol{\mathcal D}}_\mu=\frac{d{\boldsymbol{\mathcal{L}}}_\mu} {d|\mu|}\cdot\left[\frac{d(\vD_\mu/\kappa)}{d|\mu|}\right]^{-1}= 
\frac{2}{|\mu|\cF_d}\approx\frac{2}{\alpha_{|\mu|}}.
\lab{dos}
\ee
Thus DoS is inversely proportional to the residual finesse.
If DoS is close to and less than $1$,  then the residual spectrum is sparse. 

\begin{figure}[p!t]
	\centering{	
		\includegraphics[width=0.49\textwidth]{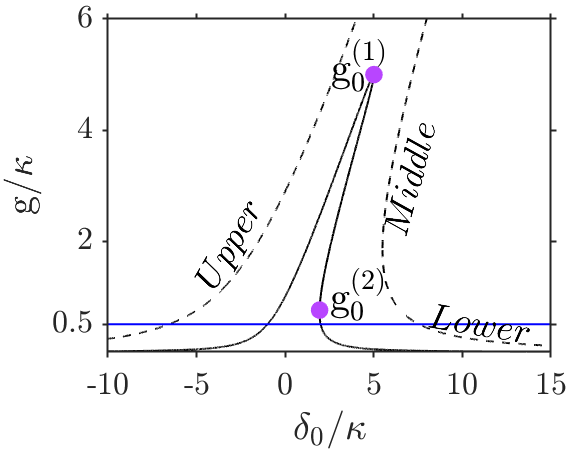}	
	}
	\caption{ Normalised intra-resonator cw-state power computed from 
		Eq.~\bref{cw}: 	$\cW/\cW_*=5$ (full  line), $\cW/\cW_*=100$ (dashed  line). 
		Horizontal blue line shows the FWM threshold as per Eq.~\bref{mn2}.  Magenta points are given by  
		Eqs.~\bref{gg} for $\mu=0$.  $\vg/\gamma$  equals the intra-resonator cw power  in Watts, see Table I.}
	\label{f1}
\end{figure}

\subsection{Power scaling}
If the power coupling efficiency is $\eta=\kappa_c/\kappa<1$, and $\kappa_c$ 
is the coupling loss, then the laser power, $\cW$,  
that couples to a ring resonator is calculated as \cite{book1}
\be
\cW=\frac{\pi}{\eta\cF}\cH^2.
\lab{pp}
\ee
We introduce here the characteristic power scales $\cW_*$ and $\kappa/\gamma$ for the 'on-chip' laser power
and for the intra-resonator power, respectively,
\be
\cW_*=\frac{\pi}{\eta}\frac{\kappa^2}{D_1\gamma},~\frac{\cW}{\cW_*}=\frac{\cH^2}{\kappa/\gamma}.
\lab{ps}
\ee
We note, that $D_1$ is inversely proportional to the resonator length and $\gamma\sim 1/S$, hence $\cW_*\sim V$, where $V$ is the mode volume. Also, $\kappa=\om_0/Q$,
where $Q$ is the quality factor. Hence, $\cW_*\sim V/Q^2$, which is in the agreement with the laser power FWM threshold scaling previously used and derived in connection with the microresonator 
FWM~\cite{revv2,f1,f2,f3,f4}. 

Appendix provides detailed estimates for the parameters, while 
Table I sums up the quality factors, linewidths, finesse dispersions, and power scaling factors for typical microresonator examples.
All our results and data are given in the transparently scaled form, so that a reader can either customise the scales if needed, or see directly how and if the numbers in Table I fit with the characteristics of the available or projected samples of their interest. 

\begin{table}
\caption{\label{tab1}Parameters for the bulk crystalline CaF$_2$ 
	and integrated Si$_3$N$_4$ resonators, see Appendix and text for further details. $Q$ is quality factor, $\kappa$ is linewidth, $\cF_d=D_2/\kappa$ is finesse dispersion, $\cW_*$ is the laser pump power scale, and $\kappa/\gamma$
is the intra-resonator power scale.}
	\begin{ruledtabular}
		\begin{tabular}{lllll}
$Q$& $\kappa/2\pi$ & $|\cF_d|$&~~$\cW_*$&~$\kappa/\gamma$
			\\ \hline\hline
			~&~&{\bf CaF}$\mathbf{_2}$&~&~\\ \hline\\
			$10^9$&$200$~kHz &$0.005$& $1$~mW& $20$~W\\
			$10^{11}$&$2$~kHz &$0.5$& $0.1~\mu$W& $0.2$~W\\
\hline\hline		~&~&{\bf Si}$\mathbf{_3}${\bf N}$\mathbf{_4}$&~&~
\\ \hline\\
$10^6$&$100$~MHz &$0.5$& $4$~mW& $5$~W\\
$10^{7}$&$10$~MHz &$5$& $40~\mu$W& $0.5$~W
		\end{tabular}
	\end{ruledtabular}
\end{table}

\subsection{CW-state and its stability}
We define the intra-resonator power of the single-mode  cw-solution of Eq.~\bref{ll}  
via the identity $|\psi|^2=\vg/\gamma $, where the 
newly introduced parameter $\vg\ge 0$ has units of frequency.
Then, the complex cw, i.e., continuous wave, amplitude is 
\be
\psi=\sqrt{\frac{\vg}{\gamma}}~e^{i\phi_0}
=\frac{-i\tfrac{1}{2}\kappa\cH}{\delta_0-\vg-i\tfrac{1}{2}\kappa},
\lab{cw0}\ee
where $\phi_0=\arg\psi_0$.
Taking absolute value of Eq. \bref{cw0} yields the bistability equation,
\be
\frac{\vg}{\kappa}+\frac{4\vg}{\kappa}
\left(\frac{\delta_0}{\kappa}-\frac{\vg}{\kappa}\right)^2=\frac{\gamma\cH^2}{\kappa}=\frac{\cW}{\cW_*},
\lab{cw}
\ee
see Fig. \ref{f1}. In order to estimate the intra-resonator cw power, $\vg/\gamma$, the dimensionless 
$\vg/\kappa$ parameter shown in 
Fig.~\ref{f1} and used through the rest of the text and figures should be multiplied by $\kappa/\gamma$, 
see Table I for some typical values.

Perturbing the cw with a pair of sideband modes~\cite{llprl}, 
\be
\psi=\sq{\vg/\gamma}~e^{i\phi_0}
+\psi_{\mu}e^{i\mu\ta}+\psi^*_{-\mu}e^{-i\mu\ta},
\lab{ser}\ee
and setting 
$\psi_{\pm\mu}\sim e^{\lambda_\mu t}$ gives
\begin{equation}
i\lambda_\mu\begin{bmatrix}
\psi_\mu \\ \psi_{-\mu}
\end{bmatrix}=\begin{bmatrix}
\vD_\mu-2\vg-i\tfrac{1}{2}\kappa & -\vg e^{i2\phi_0}  \\
\vg  e^{-i2\phi_0} & -\vD_\mu+2\vg-i\tfrac{1}{2}\kappa 
\end{bmatrix}\begin{bmatrix}
\psi_\mu \\ \psi_{-\mu}
\end{bmatrix},
\lab{mat}
\end{equation}
and hence 
\begin{align}
\lambda_\mu\big(\lambda_\mu+\kappa\big)&=
3\big(\Omeg^{(1)}_{\mu}-\Omeg\big)\big(\Omeg-\Omeg^{(2)}_{\mu}\big)\lab{lam}\\
&=3\big(\vD_\mu-\vg\big)\big(\vg-\tfrac{1}{3}\vD_\mu\big)-\tfrac{1}{4}\kappa^2,
\nn 
\end{align}
where
\bsub
\lab{gg}
\begin{align}
&\frac{\vg^{(1)}_{\mu}}{\kappa}=\frac{2}{3}\frac{\vD_\mu}{\kappa}+\frac{1}{3}
\sqrt{\frac{\vD_\mu^2}{\kappa^2}-\frac{3}{4}},
\lab{gg1}\\
&\frac{\vg^{(2)}_{\mu}}{\kappa}=\frac{2}{3}\frac{\vD_\mu}{\kappa}-\frac{1}{3}
\sqrt{\frac{\vD_\mu^2}{\kappa^2}-\frac{3}{4}}.
\lab{gg2}
\end{align}
\esub
$\vg$ couples the $\psi_\mu$ and $\psi_{-\mu}$ sidebands, cf., Refs. \cite{f4,arnold}.  
The coupling is anti-Hermitian, and therefore it provides gain and seeds the instability.

Eq.~\bref{lam} has two roots, 
\be
\lambda^{\pm}_\mu=-\tfrac{1}{2}\kappa\pm\sqrt{3(\vD_\mu-\vg)(\vg-\tfrac{1}{3}\vD_\mu)}.
 \lab{roots}
\ee
The real part of  $\lambda^{-}_\mu$ is unconditionally negative, while the other root yields
the threshold condition, Re$\lambda_\mu^+=\lambda_\mu^+=0$,
\be
\vg=\vg^{(1)}_\mu, \vg=\vg^{(2)}_\mu, 
\lab{e7}
\ee
with   
\be\vg^{(2)}_\mu<\vg<\vg^{(1)}_\mu
\lab{e6}
\ee being  an 
interval of $\vg$  with the exponential growth of the sideband powers.
Thus, $\vg_\mu^{(i)}\ge 0$, $i=1,2$, are  the critical values 
of the $\pm\mu$ sideband coupling  limiting the respective FWM gain range from below, $i=2$, and from above, $i=1$. Eq. \bref{e6} can be compared with the cw instability interval reported in Ref.~\cite{llprl} and using $\tfrac{1}{2}\mu^2\cF_d$ as a control parameter.

Two lines, $\vg=\vg_\mu^{(i)}$, merge at $\vg_\mu^{(1)}=\vg_\mu^{(2)}$, 
and extend for
\be
\frac{\vD_\mu}{\kappa}\ge\frac{\sqrt{3}}{2}
\lab{e8}
\ee
through the $|\mu|$-instability range.
The value of $\vg$ corresponding to the equal sign in Eq.~\bref{e8} is
\be
\frac{\vg}{\kappa}=\frac{1}{\sqrt{3}}=\frac{\vg_\mu^{(1)}}{\kappa}=\frac{\vg_\mu^{(2)}}{\kappa}.
\lab{e8a}
\ee

The minimum of $\vg_\mu^{(2)}$ in $\delta_0$ corresponds to
the lowest intra-resonator power, $\vg_\mu^{(2)}/\gamma$, required to kick-start FWM for a given $\pm\mu$ pair, 
i.e., to the {\em FWM threshold}. Coordinates of the minimum are 
\bsub
\lab{mn}
\begin{align}
& \frac{\vD_\mu}{\kappa}=1,\lab{mn1}\\
& \frac{\vg}{\kappa}=\frac{1}{2}=\min_{\delta_0}\frac{\vg_\mu^{(2)}}{\kappa}.
\lab{mn2}
\end{align}
\esub
For $\kappa=2$, Eqs.~\bref{mn} match  equations for $a(n_c)$ and  $E_{s}$ from Ref.~\cite{llprl}. 

The $\vg_\mu^{(2)}$ values at their minima are $\mu$ independent, while recovering
$\delta_0$ at the same points from Eq.~\bref{mn1} and Eq.~\bref{resa} gives
\be
\frac{\delta_0^{(\mu F)}}{\kappa}=1-\frac{1}{2}\mu^2\cF_d.
\lab{df}
\ee
Let us note, that stability analysis in this  subsection did not touch on the problem of the interplay between the sidebands with different $|\mu|$, 
and therefore did not require using the residual finesse, $\al_\mu$.
All this becomes central in what follows.

\begin{figure*}[p!t]
	\centering{
		\includegraphics[width=0.9\textwidth]{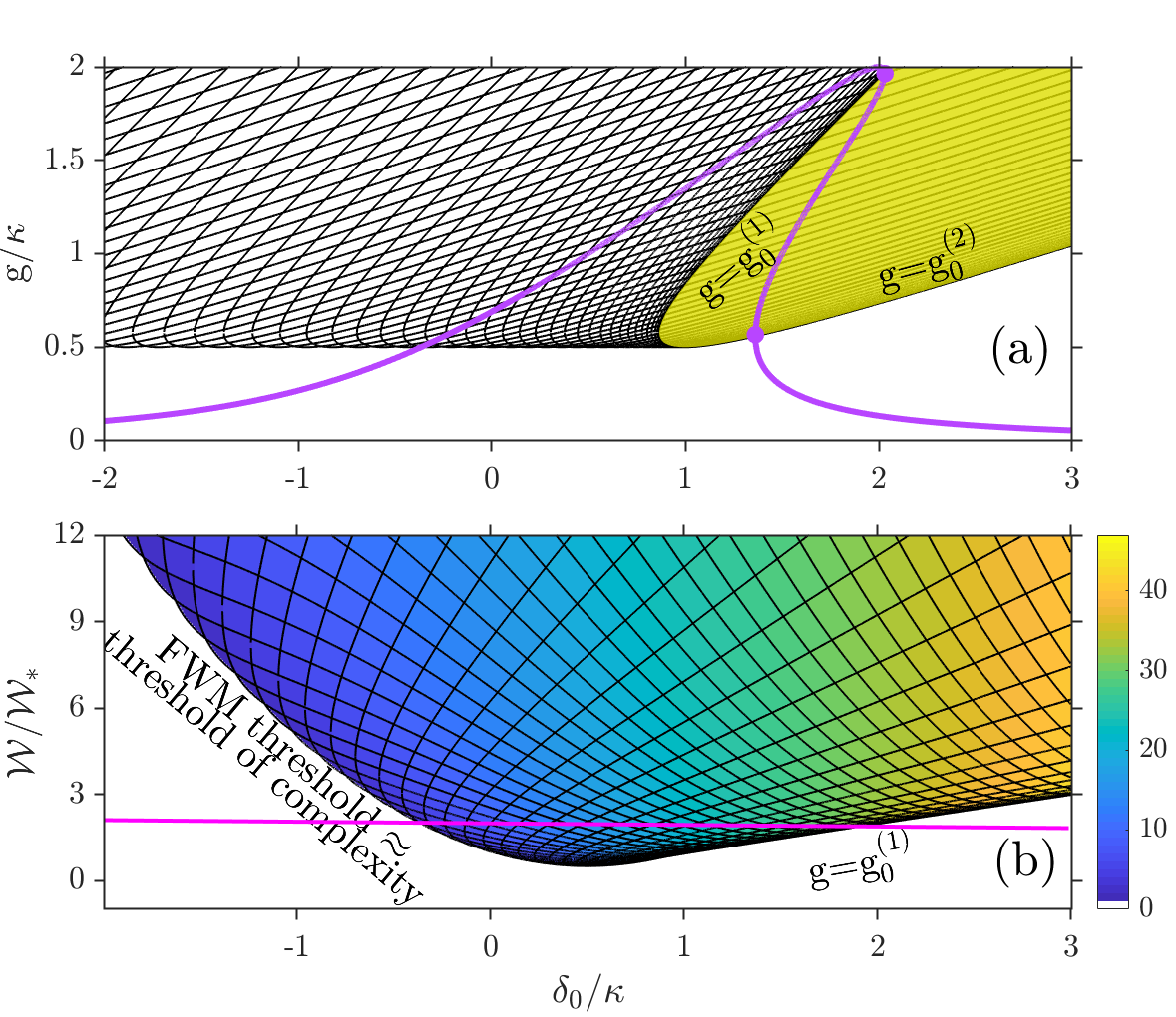}	
	}
	\caption{An illustration of how the quasi-continuous residual spectrum, $|\cF_d|=|D_2|/\kappa\ll 1$, leads to the coinciding  FWM and  complexity thresholds. 
		{\bf (a)} $\vg/\kappa=\vg_\mu^{(i)}/\kappa$
		vs $\delta_0/\kappa$ for $\mu=0,1,\dots 64$ and   $\cF_d=0.005$ (anomalous dispersion). This is an example of the quasi-continuous residual spectrum forming the quasi-smooth line converging to $\vg/\kappa=1/2$, see the last column in Table I for scaling of $\vg/\kappa$ to the intra-resonator  cw power, $\vg/\gamma$.  Yellow shading shows the range  corresponding to the middle branch of the cw-state. 
		Magenta line  is the cw-state for  $\cW/\cW_*=2$.
		{\bf (b)} Normalised threshold laser power, see Eq. \bref{physw},  vs $\delta_0/\kappa$ computed using  the $\vg_\mu^{(i)}$  data from (a) taken outside the yellow shading, so that the plot applies to the upper branch of the cw-resonance. The colorbar shows the number of the simultaneously growing sideband pairs. Magenta line is drawn at $\cW/\cW_*=2$. }
	\label{f2}
\end{figure*}

\begin{figure*}[p!t]
	\centering{
		\includegraphics[width=0.85\textwidth]{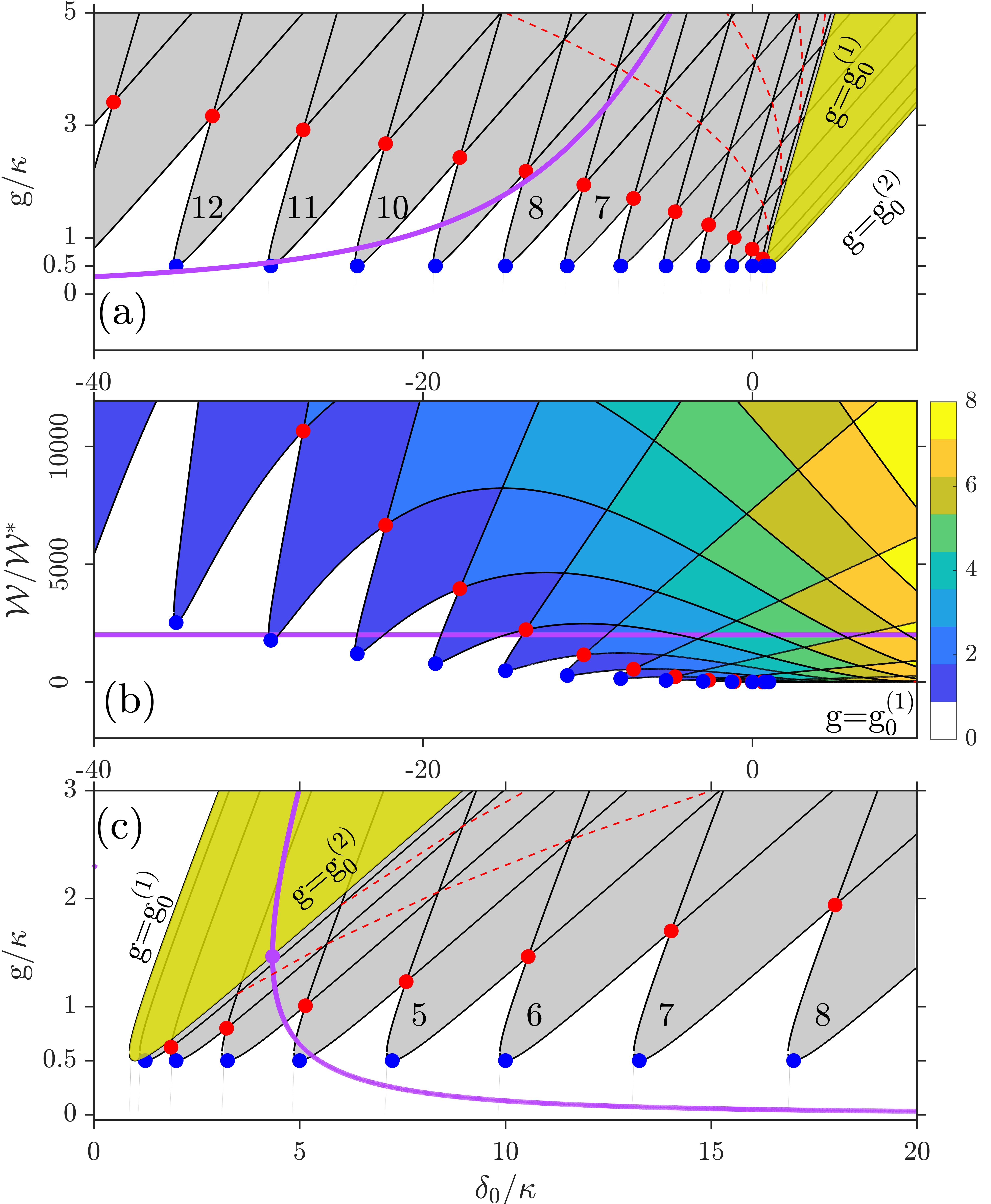}	
	}
	\caption{An illustration of how the sparse residual spectrum, $|\cF_d|=|D_2|/\kappa\sim 1$, leads to the very different  FWM  (blue points) and  complexity (red points) thresholds. 
		{\bf (a)}~$\vg=\vg_\mu^{(i)}$
		vs $\delta_0$ for $\mu=0,1,\dots 13$ and   $\cF_d=0.5$ (anomalous  dispersion). Yellow shading shows the range of parameters corresponding to the middle branch of the cw-state, $\vg_0^{(2)}<\vg<\vg_0^{(1)}$. Grey shading shows the range of $\mu\ne 0$ instabilities.
		Magenta line shows the $\delta_0<0$ tail of the upper  branch of the cw-state for  $\cW/\cW_*=2000$. The in-figure numbers show $|\mu|$ for the respective tongues. For the chosen  $\cW/\cW_*$, the $|\mu|=9,10,11$ tongues make isolated in $\delta_0$ instability intervals. One sideband pair, $\pm\mu$, is unstable inside the quasi-rhombs starting at the blue points, i.e., FWM threshold. Two pairs, $\pm\mu$, $\pm(\mu+1)$, are simultaneously unstable inside the quasi-rhombs extending upwards from the red points, i.e., threshold of complexity. 
		{\bf (b)}~Normalised threshold laser power, see Eq. \bref{physw},  vs $\delta_0/\kappa$ computed using  the $\vg_\mu^{(i)}$ data from (a) taken outside the yellow shading. The colorbar shows the number of the simultaneously growing sideband pairs.	Magenta line is drawn at $\cW/\cW_*=2000$. {\bf (c)}~is like (a), but for   $\cF_d=-0.5$ (normal  dispersion). Magenta line shows the lower and middle branches of the cw-state,  $\cW/\cW_*=50$. The $|\mu|=4$ tongue makes an isolated  instability interval around  $\delta_0/\kappa\approx 5$.
	 For scaling of $\vg/\kappa$ in (a) and (c) to the intra-resonator cw power $\vg/\gamma$, see the last column in Table I.}
	\label{f3}
\end{figure*}

\section{Instabilities, tongues and thresholds}
Insight into the intersections, which presence or absence is also not immediately obvious, of the stability boundaries  given by Eqs. \bref{e6} for different $\mu$ in the space of physical parameters reveals how a resonator transits from the stable cw-state to the sideband generation~\cite{arnold}. 
To investigate how the transition from the quasi-continuous, $|\alpha_\mu|\ll 1$, to sparse, $|\alpha_\mu|\gg 1$, residual spectra  impacts the net instability threshold one can plot the threshold laser power, $\cW$,  vs  $\delta_0$  for every $\mu\in\mathbb{Z}$ and different $\cF_d$, see Figs. \ref{f2}, \ref{f3}. 
The respective powers are calculated by substituting $\vg_\mu^{(i)}$ into Eq.~\bref{cw} 
and using Eqs.~\bref{pp}, \bref{ps}, 
\be
\frac{\cW_\mu^{(i)}}{\cW_*}=\frac{\vg_\mu^{(i)}}{\kappa}+\frac{4\vg_\mu^{(i)}}{\kappa}
\left(\frac{\delta_0}{\kappa}-\frac{\vg_\mu^{(i)}}{\kappa}\right)^2,
\lab{physw}
\ee
Eq. \bref{physw} is dimensionless and parameterised  by the finesse dispersion, $\cF_d$,  and, by the normalised laser detuning, $\delta_0/\kappa$.  

\begin{figure*}[p!t]
	\centering{	
		\includegraphics[width=0.9\textwidth]{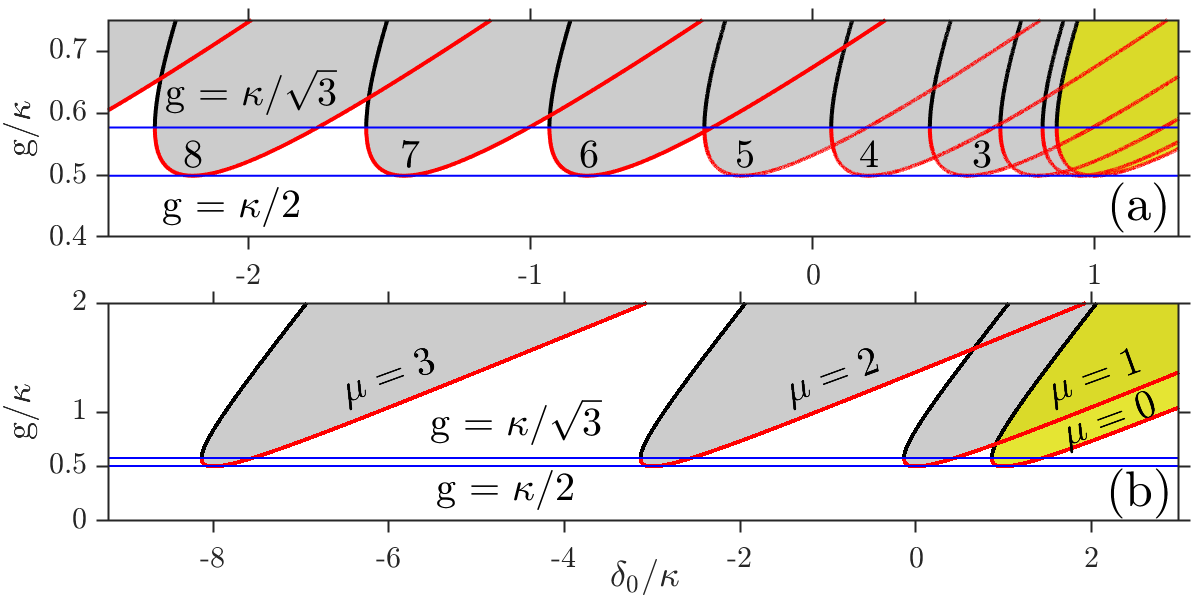}	
	}
	\caption{{\bf (a)} $\cF_d=0.1$: Transition from the red-red crossing of the $\mu$ and $\mu+1$ instability lines to the red-black ones, $\wh\mu\approx 5.3$, $|\wh\delta_0/\kappa|\approx 0.4$, see section IV.C. The last red-red crossing is $\mu=5$ to $\mu=6$.  
	{\bf (b)} $\cF_d=1.5$: In this case, all the crossings are the red-black ones, since $\cF_d>\wt\cF_d$. Blue lines show the minima of the tongues, i.e., FWM threshold, $\vg/\kappa=1/2$,  and the 	$\vg_\mu^{(1)}=\vg_\mu^{(2)}$ points, where $\vg/\kappa=1/\sq{3}$, i.e., bistability threshold. Black and red lines correspond to $\textsl{g}_\mu^{(1)}$ and $\textsl{g}_\mu^{(2)}$, respectively.}
	\label{f4}
\end{figure*}

To facilitate with the theory developed below it is important to analyse the   $\vg=\vg_\mu^{(i)}$ vs $\delta_0$ plots, see Figs. \ref{f2}-\ref{f4}.
For $|\cF_d|\ll 1$ the common envelope of the threshold lines flattens and approaches  the $\vg/\kappa=1/2$ line, see Fig.~\ref{f2}(a).  Instead, with $|\cF_d|$ increasing and approaching one, the instability tongues are shaping up with the gradually increasing power (vertical) contrast and  frequency (horizontal) sparsity, see Fig.~\ref{f3} and Ref. \cite{arnold}. The Arnold-tongue structure implies  that small changes of the pump frequency can lead to the large variations in the power threshold, cf., Figs.~\ref{f2}(b) and \ref{f3}(b). The $|\mu|\ne 0$ tongues unfold either left or right from  the $\mu=0$ one, depending on the sign of $\cF_d$, cf.,  Fig.~\ref{f3}(c) for normal dispersion and, e.g., Fig.~\ref{f3}(a) for anomalous.

In the nonlinear dynamics, Arnold tongues are classified by their synchronisation order
$n~:~m$, where $n$ is the number of nonlinear pulses coming through the system per  $m$ periods of the linear oscillator~\cite{bio}. In our case, the linear round trip time is $2\pi/D_1$ and periods of the synchronised waveforms
given Eqs.~\bref{ser}, \bref{llb} are $2\pi/|\mu|D_1$, thus the corresponding orders are $|\mu|~:~1$.
Practically, mLLE Arnold tongues  are defined as per  the convergence of 
their minima to the points of resonances, $\vD_\mu=0$, in the linear residual spectrum, i.e., 
for $\vg,\kappa\to 0$, \cite{arnold,bio}. Only the dispersive tails of the cw-resonance, see Fig. \ref{f1}, withstand the $\vg\to 0$ limit.
While $\vD_\mu=0$ opens up as $\delta_0=-\tfrac{1}{2}\mu^2D_2$, and hence the linear resonances occur for $D_2>0$ at $\delta_0<0$, i.e., along the tail of the upper branch, and for $D_2<0$ at $\delta_0>0$, i.e., along the lower branch, see Fig.~\ref{f1}.

The $\vg^{(2)}_0<\vg<\vg^{(1)}_0$ condition 
corresponding to the yellow shaded regions in Figs.~\ref{f2}(a), and \ref{f3}(a),(c)
is an important reference condition.
If one tunes the input power, $\cW$, then the tip of the nonlinear cw-resonance, i.e., the point where the upper and middle branches connect, slides along the $\vg=\vg^{(1)}_0$ line. Simultaneously, the turning point connecting the middle and lower branches slides along the $\vg=\vg_0^{(2)}$. Hence, the cw-solution can either go through the parameter ranges providing uninterrupted FWM gain, or  cut across one or more isolated Arnold tongues, so that the instability and stability intervals intermit, see 
and compare Figs.~\ref{f2}(a), \ref{f3}(a),(c).  As soon as the $\vg/\kappa$  
vs $\delta_0$ cw tail drops below $1/2$, then the FWM instability becomes prohibited, as per  Eqs.~\bref{mn}. 
Nonlinear transformation $\vg\to\cW$, Eq.~\bref{cw}, maps the magenta coloured bistable cw-states in 
Figs.~\ref{f2}(a), \ref{f3}(a) to the horizontal lines corresponding to the constant power levels in the $(\delta_0,\cW)$ plots, Figs.~\ref{f2}(b), \ref{f3}(b).

The yellow shading in Figs.~\ref{f2}(a), \ref{f3}(a,c) embraces all and only parameter values belonging to the middle branch of the bistability loop. 
$\delta_0$ point where the middle and upper branches merge is
slightly shifted from the $\text{max}_{\delta_0}(\vg)$ point, see Fig.~\ref{f5}(b).
The instability tongues that unfold to the right for $\cF_d<0$ in Fig.~\ref{f3}(c) belong to the lower branch of the cw-sate in Fig.~\ref{f1}. Similarly, 
the tongues that unfold to the left for $\cF_d>0$ in Figs.~\ref{f2}(a), \ref{f3}(a) belong to the upper branch.

Eqs.~\bref{gg}, \bref{physw} divide the $(\delta_0,\vg)$ and $(\delta_0,\cW)$ parameter spaces into the well defined sections  corresponding to the simultaneous growth of one, two, three and more pairs of modes, see Figs.~\ref{f2}(b), \ref{f3}(b). In the $|\cF_d|\ll 1$ case, see Fig.~\ref{f2}(b), these sections form a quasi-continuous pattern with the density of states increasing with $\vg$. The areas immediately above the threshold line for the negative and positive $\delta$'s are however structured differently.
For $\delta<0$, the segments along the instability boundaries contain only one unstable $\pm\mu$ pairs, while for the different $|\mu|$ accomulate together for $\delta>0$.
 
If $|\cF_d|$ is brought close to $1$, see Fig. \ref{f3}(b), then the density of the quasi-rhombs becomes much lower. Simultaneously, the  parameter space, across the range of the relatively small pump powers and for $\delta<\kappa$,
becomes divided into $|\mu|$-specific instability tongues separated by the stability intervals.  
Crossing into the first layer of rhombs in Fig.~\ref{f3}(b), i.e., into the rhombs  having
their bottom vertices marked with the blue points, implies instability of only one $\pm\mu$ sideband pair. Crossing into the second layer of rhombs having their bottom vertices marked with the red points implies the simultaneous instability relative to the $\pm\mu$   and  $\pm(\mu\pm 1)$ sideband pairs, and so on.

The  blue dots at the minima of the tongues correspond to the conditions in Eq.~\bref{mn} and they make a threshold that is referred here as the FWM threshold.
Reaching the FWM threshold implies exponential amplification of the sideband pairs, but the pump frequency needs to be tuned carefully to trigger frequency conversion into a particular $\pm\mu$ pair. When the second layer of rhombs is reached, see red dots, only then 
the stability gaps between the tongues disappear and the simultaneous 
instabilities of the $|\mu|$ and $|\mu\pm 1|$ sidebands become possible.
Note, that the red dots are the cusps, and are not differentiable in $\delta_0$, unlike the blue dots found from the $\p_{\delta_0}\vg_\mu^{(2)}=0$ condition, 
Eq.~\bref{mn}. The line of the red dots  is referred by us as the {\em threshold of complexity}~\cite{arnold}.
This is the first and only in the sequence of the higher-order thresholds that touches the intervals of the cw-state stability. 
$\cW_*$ provides quite a small scale, see Table I. So that, fixing $\cW/\cW_*=5000$, the two ultrahigh-Q cases, see Fig.~\ref{f3}, give $\cW=0.5$mW (CaF$_2$) and $200$mW (Si$_3$N$_4$) of the on-chip laser power, and only few Watts of the intra-resonator powers. So that, the tongue structure and complexity threshold are within a practical rich even in cases of smaller $Q$'s. 

Section IV accomplishes formulating formal conditions  for the first and the higher-order lines of the cusp points and solves these conditions for $\vD_\mu$, $\delta_0$. $\vg$ and power values  along the threshold of complexity are derived in Section V.

Before moving on, let us remark, that the external contour of the power plot 
for $|\cF_d|\ll 1$, see Fig.~\ref{f2}(b), should be compared with the same thresholds derived and discussed in Ref.~\cite{chembo1}. 
The equation for the contour in Fig.~\ref{f2}(b) is found by applying $\vg^{(i)}_\mu/\kappa=1/2$ in Eq.~\bref{physw}, 
\be\frac{\cW}{\cW_*}=\frac{1}{2}+2
\left(\frac{\delta_0}{\kappa}-\frac{1}{2}\right)^2,\lab{con}\ee
and should be used for  $\delta_0<\delta_0^{(0F)}=\kappa$, $\cF_d>0$. Substitution of $\vg=\vg_0^{(1)}$ in Eq.~\bref{physw} follows the contour in Fig.~\ref{f2}(b) for $\delta_0>\kappa$. Eq.~\bref{con} also applies for $\delta_0>\delta_0^{(0F)}$, but if one takes  the lower cw branch and $\cF_d<0$.

\section{Threshold of complexity: Detunings at the cusps}
We now focus on the development of the analytical theory for the threshold of complexity. The $|\mu|$ specific instability boundaries consist in general from the two different conditions, $\vg=\vg_\mu^{(1)}$ and $\vg=\vg_\mu^{(2)}$. In order to look into 
how these two conditions interact along the threshold of complexity, we make two new $\vg$ vs $\delta_0$ plots, see  Fig.~\ref{f4}, and assign the black color to the $\vg=\vg_\mu^{(1)}$  and the red one to $\vg=\vg_\mu^{(2)}$ conditions, respectively.  We  note, that 
\be
\vD_\mu=\vD_{-\mu},~\vg_\mu^{(i)}=\vg_{-\mu}^{(i)},
\lab{mu1}
\ee
 and hence fixing 
$\mu\ge 0$
does not restrict generality of the theory developed below.

\subsection{Red-red intersections}
For the relatively small  $|\cF_d|$, there exists
a range of  $\delta_0$ where the complexity threshold is made
by the intersections of the $\vg=\vg_\mu^{(2)}$ and $\vg=\vg_{\mu+ 1}^{(2)}$ lines, i.e., by 
the red-red intersections, see Fig.~\ref{f4}(a). The range of  
$|\cF_d|$ providing the red-red intersections is limited from above $|\cF_d|<|\wt\cF_d|$, where $\wt\cF_d$ is worked out later, see Eq.~\bref{fd}.
For either sign of $\cF_d$  the coordinates of the 
red-red intersections are found by solving
\be  
\vg=\vg_{\mu}^{(2)}=\vg_{\mu+1}^{(2)},  \lab{eq12}
\ee
Applying  Eqs. \bref{gg}, \bref{eq12} we find
\begin{align}
&|\cF_d|<|\wt\cF_d|,\nn \\ 
&\frac{2\vD_\mu}{\kappa}-
	\sqrt{\frac{\vD_\mu^2}{\kappa^2}-\frac{3}{4}}=
	\frac{2\vD_{\mu+1}}{\kappa}-
	\sqrt{\frac{\vD_{\mu+1}^2}{\kappa^2}-\frac{3}{4}}.
	\lab{e13}
	\end{align}
Introducing the residual finesse, as given	by Eq.~\bref{rfa}, to link $\vD_{\mu+1}$
and $\vD_\mu$ gives
 \begin{align} 
 0=&2\alpha_\mu+\sqrt{\frac{\vD_\mu^2}{\kappa^2}-\frac{3}{4}}-
\sqrt{\left(\frac{\vD_{\mu}}{\kappa}+\alpha_\mu\right)^2-\frac{3}{4}}
\lab{e16}\\
= & \cR_{22}(\alpha_\mu,\vD_\mu).\nn
\end{align}
Here, the subscript '$22$', reminds  that Eq.~\bref{e16}
corresponds to the intersections between $\vg^{(2)}_\mu$ and $\vg^{(2)}_{\mu+1}$. 

Squaring  Eq.~\bref{e16}   twice yields an elegant equation for $\vD_\mu$,
\be
\left(\frac{\vD_\mu}{\kappa}+\frac{\alpha_\mu}{2}\right)^2=1+\alpha_\mu^2.
\label{g11}
\ee
Recalling Eq.~\bref{e8} selects the positive  root of Eq.~\bref{g11}, when it is solved for $\vD_\mu$, 
\be
\frac{\vD_\mu}{\kappa}=\sq{1+\alpha_\mu^2}-\frac{\alpha_\mu}{2}.
\lab{e18a}
\ee
Inserting Eq.~\bref{e18a} back into Eq.~\bref{e16} makes a function of a single variable $\al_\mu$, i.e.,
$\cR_{22}(\al_\mu,\vD_\mu(\al_\mu))$. $\cR_{22}$ is zero  for 
\be
-|\al_{\wh\mu}|<\al_\mu<|\al_{\wh\mu}|, 
\lab{ac1}
\ee
where $\al_{\wh\mu}$ is a constant, see Fig.~\ref{f5}(a) and Eq.~\bref{e22a}.

Recovering $\delta_0$  from Eqs.~\bref{e18a}, \bref{resa} yields
\be
\frac{\delta_0^{(\mu C)}}{\kappa}=\sq{1+\al_\mu^2}-\frac{\al_\mu}{2}-\frac{1}{2}\mu^2\cF_d,
\lab{dc}
\ee
that holds at the red points in Figs.~\ref{f1}(b), \ref{f3}(a), cf., 
Eq.~\bref{df} valid at the blue points.

Eq.~\bref{e18a} is derived again in the next subsection, where we point at  its convergence to Eq.~\bref{mn1} in the limit $\al_\mu\to 0$, and proceed with further important implications derived from this critical condition.

\subsection{Red-black intersections}
As $|\mu|$ is increased, the red-red crossings along the universal threshold are replaced with the red-black ones for  $\mu>\wh\mu$, see Fig.~\ref{f4}(a). If $|\cF_d|$ is large enough,
i.e., $|\cF_d|>|\wt\cF_d|$, then the intersections can become exclusively of the red-black type, including the $\mu=0$ and $\mu=1$ one, see Fig.~\ref{f4}(b).  Both $\wh\mu$ and $\wt\cF_d$ are worked out below, see Eqs. \bref{e26}, \bref{fd}.

The red-black intersections for $\cF_d>0$ and $\cF_d<0$ are given by different conditions,
\bsub    
\lab{g2}
\begin{align}
&\cF_d>0:~~ \vg=\vg_{\mu}^{(1)}=\vg_{\mu+1}^{(2)}, 
\lab{co3}\\
&\cF_d<0:~~ \vg=\vg_{\mu}^{(2)}=\vg_{\mu+1}^{(1)}.
\lab{co4}
\end{align}
\esub
A reason for the two conditions here and one in the previous subsection is that now 
$\vg_\mu^{(1)}$ is matched with $\vg_{\mu+1}^{(2)}$, and in the previous case we matched  $\vg_\mu^{(2)}$ with $\vg_{\mu+1}^{(2)}$, so that the ordering did not matter. An accompanying factor to account for is that the tongues  tilt points in the direction of $|\mu|$ increasing for $\cF_d>0$, and  towards $|\mu|$ decreasing for $\cF_d<0$, cf., Figs.~\ref{f1}(b) and \ref{f3}(a).

\begin{figure*}[p!t]
	\centering{	
		\includegraphics[width=0.9\textwidth]{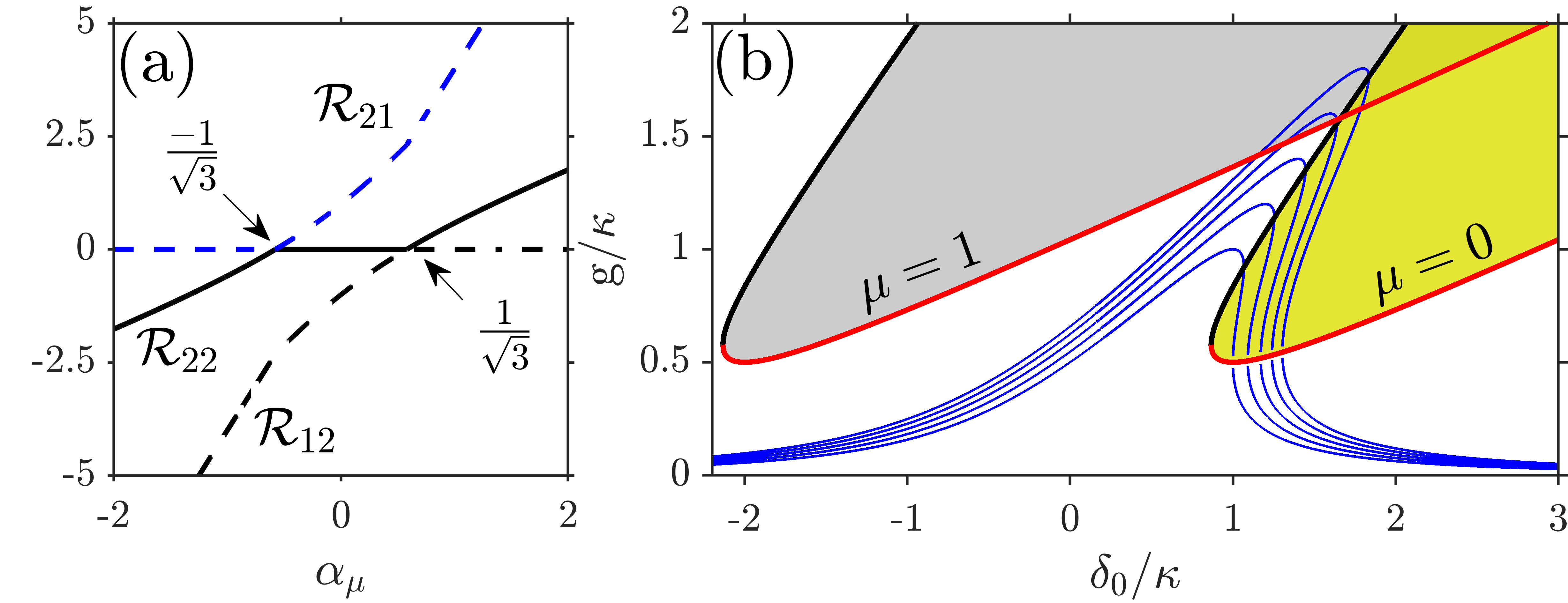}	
	}
	\caption{{\bf (a)} Functions $\cR_{jj'}$ vs residual finesse, $\al_\mu$.  $\cR_{jj'}=0$ constitute equations for the threshold of complexity. $\cR_{22}$ correspond to the red-red intersections, and $\cR_{12}$, $\cR_{21}$ to the red-black ones, see 
		Fig.~\ref{f4}. {\bf (b)} Bistability without FWM for $\cF_d=6$. First three blue lines show the bistable cw-states, when both the upper and lower branches are stable relative to the sideband excitations. Two top blue lines have the upper branch entering the $|\mu|=1$ instability range. The pump values are $\cW/\cW_*=1$, $1.2$, $1.4$, $1.6$, $1.8$ for the blue lines from bottom to top.	}
	\label{f5}
\end{figure*}

Substituting Eqs. \bref{gg} to Eqs. \bref{g2} yields, respectively,
\bsub
\lab{e20} 
\begin{align}  
&\cF_d>0:~  \frac{2\vD_{\mu}}{\kappa}+
\sqrt{\frac{\vD_{\mu}^2}{\kappa^2}-\frac{3}{4}}=
\frac{2\vD_{\mu+1}}{\kappa}-
\sqrt{\frac{\vD_{\mu+1}^2}{\kappa}-\frac{3}{4}},
\lab{e20a}
\\
&\cF_d<0:~  \frac{2\vD_{\mu}}{\kappa}-
\sqrt{\frac{\vD_{\mu}^2}{\kappa^2}-\frac{3}{4}}=
\frac{2\vD_{\mu+1}}{\kappa}+
\sqrt{\frac{\vD_{\mu+1}^2}{\kappa^2}-\frac{3}{4}}.
\lab{e20b}
\end{align}
\esub
Introducing $\alpha_\mu$, Eqs. \bref{e20} are replaced with
\bsub
\lab{e21}
\begin{align}
\nn
\cF_d>0:~ 0&=2\alpha_\mu-\sqrt{\frac{\vD_\mu^2}{\kappa^2}-\frac{3}{4}}-
\sqrt{\left(\frac{\vD_{\mu}}{\kappa}+\alpha_\mu\right)^2-\frac{3}{4}}\\
&= \cR_{12}(\alpha_\mu,\vD_\mu),
\lab{ma}\\
\nn 
\cF_d<0:~
0&=2\alpha_\mu+\sqrt{\frac{\vD_\mu^2}{\kappa^2}-\frac{3}{4}}+
\sqrt{\left(\frac{\vD_{\mu}}{\kappa}+\alpha_\mu\right)^2-\frac{3}{4}}\\
&=\cR_{21}(\alpha_\mu,\vD_\mu).
\lab{mb}
\end{align}
\esub
Resolving Eqs.~\bref{e21} for $\vD_\mu$ also gives Eq.~\bref{e18a}.
Eq.~\bref{ma} and Eq.~\bref{mb} are satisfied for $\al_\mu>|\al_{\wh\mu}|$
and   $\al_\mu<-|\al_{\wh\mu}|$, respectively, see Fig.~\ref{f5}(a).

{\em Thus, Eq.~\bref{e18a} specifies $\vD_\mu$, and hence $\delta_0$, at the  cusp points  along all
of the threshold of complexity
and should be compared with Eq.~\bref{mn1} valid at the tongue minima along the FWM threshold. }

In the limit $\al_\mu\to 0$, Eq.~\bref{e18a} transforms to Eq.~\bref{df} valid at the FWM threshold. Noting that $\vg_\mu^{(i)}$  are expressed via $\vD_\mu$, we can claim that we have demonstrated analytically that the first line of cusps, i.e., the threshold of complexity, 
and the line of minima of $\vg_\mu^{(2)}$, i.e., the FWM threshold, converge to a single line in the limit of vanishing residual finesse, see Fig.~\ref{f2}.

The threshold where three pairs of side-bands become simultaneously unstable for $\cF_d>0$ is found from the condition $\vg_{\mu}^{(1)}=\vg_{\mu+2}^{(2)}$, and 
$\vD_{\mu+1}-\vD_\mu=\al_\mu$ in the above calculations should be replaced with $\vD_{\mu+2}-\vD_\mu=\al_\mu+\al_{\mu+1}=2(\alpha_\mu+\tfrac{1}{2}\cF_d)$. In general, a condition for $N\ge 2$ sidebands pairs, i.e., $\pm\mu$, $\pm(\mu+1)$, $\dots,$ $\pm(\mu+N-1)$, to become unstable  is
$\vg_{\mu}^{(1)}=\vg_{\mu+N-1}^{(2)}$. $\vD_\mu$ at the respective cusp points is
found applying the substitution $\al_\mu\to \al_\mu^{(N)}$ in Eq. \bref{e18a}, where
\begin{align}
\al_\mu^{(N)}=&(N-1)\left(\alpha_\mu+(N-2)\frac{\cF_d}{2}\right),
\lab{npair}
\\ \nn & N=2,3,\dots;~\al_\mu=\al_\mu^{(2)}.
\end{align}
Again, this  applies for both the red-red and red-black intersections. 
Thin dashed red lines in Figs.~\bref{f3}(a),(c) connect the cusp point along the 
$N=3$, $N=4$, and $N=5$ thresholds. We note, that the 
explicit dependence on $\cF_d$ in Eq. \bref{npair} is weak, providing $\mu\gg \tfrac{1}{2}N$. Hence, using $\al_\mu^{(N)}\approx (N-1)\al_\mu$  is a good practical approximation for the cusp points along the 
higher order thresholds.

\subsection{Crossover from the red-red to the red-black intersections:\\ Critical mode number and critical detuning}
In general, the red-red crossing make less contrasted tongues than the red-black ones. This is because the values of $\vg/\kappa$ at the red-red points have to be below
 $\vg/\kappa=1/\sqrt{3}$, see Eq. \bref{e8a}, so that they are closer to 
 the minima of the tongues at $\vg/\kappa=1/2$. While, the red-black intersections 
 can rise arbitrarily far above the minima. 
Using this, we develop a more rigid
criterion for transition from the quasi-continuous to the sparse residual spectrum, which is 
$|\al_\mu|< |\al_{\wh\mu}|$ for quasi-continuum, and $|\al_\mu|> |\al_{\wh\mu}|$ 
for the sparse residual spectrum. This approach also provides practical value for $\cF_d$, $\cF_d=\wt\cF_d$,  signalling that the  
spectrum is sparse starting from $\mu=0$.

Eqs. \bref{e16} and \bref{e21}  are  resolved by  Eq. \bref{e18a} over different 
intervals of $\al_\mu$, due to different signs in front of the  square roots in the  $\cR_{22}=0$, $\cR_{21}=0$, and $\cR_{12}=0$  equations. 
Critical values of $\alpha_\mu$, where the just mentioned intervals touch, i.e., the red-red intersections change to the red-black ones, are found imposing 
\bsub
\lab{e21q}
\begin{align}
\cF_d>0:~\cR_{22}(\alpha_\mu,\vD_\mu)&=\cR_{12}(\alpha_\mu,\vD_\mu)\Rightarrow
\frac{\vD_\mu^2}{\kappa^2}=\frac{3}{4};
\lab{e21a}
\\
\cF_d<0:~\cR_{22}(\alpha_\mu,\vD_\mu)&=\cR_{21}(\alpha_\mu,\vD_\mu)\nn
\\ &\Rightarrow
\left(\frac{\vD_\mu}{\kappa}+\al_\mu\right)^2=\frac{3}{4}.
\lab{e21b}
\end{align}
\esub

Combining Eq. \bref{e18a} and Eq. \bref{e21a} gives
\be
\cF_d>0:~ \sq{1+\alpha_{\wh\mu}^2}-\frac{1}{2}\alpha_{\wh\mu}=\frac{\sqrt{3}}{2}~ 
\Rightarrow ~
\alpha_{\wh\mu}=\frac{1}{\sqrt{3}}.\lab{e22a}
\ee
Similarly, Eq. \bref{e18a} and Eq. \bref{e21b}  lead to 
\be
\cF_d<0:~ \sq{1+\alpha_{\wh\mu}^2}+\frac{1}{2}\alpha_{\wh\mu}=\frac{\sqrt{3}}{2}~ 
\Rightarrow ~
\alpha_{\wh\mu}=\frac{-1}{\sqrt{3}}.\lab{e22b}
\ee

Recalling $\al_\mu$ definition in Eq. \bref{fna}, we find that 
\be
\mu>\wh\mu,~\text{where}~~\wh\mu=\frac{1}{\sqrt{3}|\cF_d|}-\frac{1}{2},~\wh\mu\in\mathbb{R},
\lab{e26}
\ee
corresponds to $|\al_\mu|>1/\sqrt{3}$.
This is quite a remarkable  
result that provides a resonator specific mode number  signalling transition to the Arnold tongue regime
with the threshold power sharply varying with the pump frequency.

Knowing  $\wh\mu$, 
we can estimate the pump detuning, $\wh\delta_0$, required to cross into the Arnold regime. Using Eq. \bref{mn1} gives
\begin{align}
\frac{|\delta_0|}{\kappa}\gtrsim\frac{|\wh\delta_0|}{\kappa}
=&\left|1-\tfrac{1}{2}\wh\mu^2\cF_d\right|
\lab{e28}
\\=&\left|
\frac{1}{6\cF_d}-\left(1+\frac{1}{2\sqrt{3}}-\frac{\cF_d}{8}\right)\right|.
\nn
\end{align}
The above condition can be used to check if the Arnold regime is within a reach of the pump laser tuning range in cases when a resonator sample has  $|\cF_d|\ll 1$. 
This consideration has suggested the grouping of the terms in the last part of Eq.~\bref{e28}.

\subsection{Bistability without FWM}
$\cF_d=\wt\cF_d$  marking the disappearance of the red-red crossings even from the $\mu=0$ and $\mu=1$ 
intersection for $|\cF_d|>|\wt\cF_d|$, see Fig. \ref{f4}(b), is found by taking $\wh\mu=0$, 
\be
|\wt\cF_d|=\frac{|D_2|}{\kappa}=\frac{2}{\sqrt{3}}\approx 1.155.
\lab{fd}\ee
Thus, from the point of view of Arnold tongues becoming a dominant feature of the FWM threshold, the whole of the residual spectrum starting from $\mu=0$ can be considered as sparse providing
$|\cF_d|>|\wt\cF_d|$.
Large spectral sparsity implies that the phase sensitive  FWM terms that couple the $\mu=0$ and $\mu=1$ modes oscillate on the time scales 
that are fast enough for these terms
to become negligible, so that the FWM instability become impossible. 

It means that the point of the intersection between the
$\mu=0$ and $\mu=1$ instabilities moves  above the bistability point, 
$\vg_0^{(1)}=\vg_0^{(2)}$, in the $(\delta_0,\vg)$-plane, see Fig. \ref{f5}(b).
Hence it opens up a range of pump powers and detunings that are sufficient to
make the resonator bistable, i.e., $\vg/\kappa>1/\sqrt{3}$,
but still not high enough to trigger frequency conversion.
This situation is not possible if the residual spectrum is continuous, making the frequency conversion  to start at $\vg=\kappa/2$ and as soon as $\delta_0<\delta_0^{(0F)}=\kappa$ (minimum point of $\vg=\vg_0^{(2)}$).

\section{Threshold of complexity:\\ Sideband coupling and pump power}
\subsection{Sideband coupling}
Results from the previous Section should now be used to calculate
the coupling constant $\vg$  along the threshold of complexity.
This is accomplished by substituting $\vD_\mu$ from Eq.~\bref{e18a} into 
$\vg_\mu^{(i)}$ in Eq.~\bref{eq12} and Eqs.~\bref{g2}:
\bsub
\lab{e33}
\begin{align}
&\vg=\vg_{\mu C}^{(2)}\big(\vD_\mu(\al_\mu)\big)~~\text{if}~~\al_\mu\in\Big(-\infty,\frac{1}{\sqrt{3}}\Big];\\
&\vg=\vg_{\mu C}^{(1)}\big(\vD_\mu(\al_\mu)\big)~~\text{if}~~\al_\mu\in\Big[\frac{1}{\sqrt{3}},+\infty\Big).
\end{align}
\esub
The subscripts '$C$' and '$F$' indicate that the respective quantities are taken 
along either the threshold of Complexity or along the FWM one. 
Eqs.~\bref{df} and \bref{dc} used the same superscripts.

\begin{figure*}[p!t]
	\centering{	
		\includegraphics[width=0.8\textwidth]{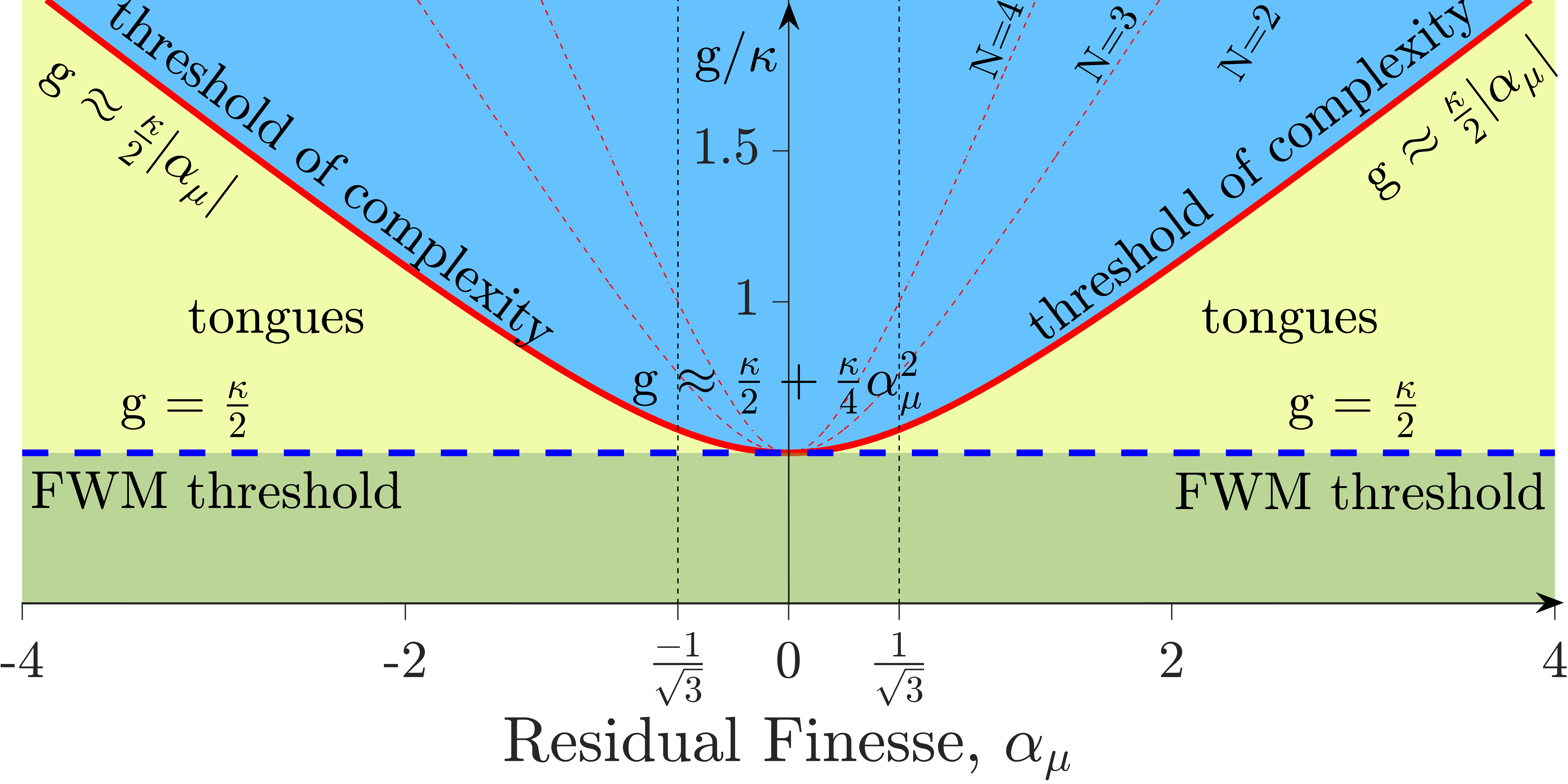}	
	}
	\caption{FWM (blue dashed line) and complexity (red full line) thresholds  in the  sideband 
		coupling strength, $\vg$, and the residual finesse, $\al_\mu\approx\mu\cF_d$, parameter space. 
		For scaling of $\vg/\kappa$  to the intra-resonator cw power $\vg/\gamma$, see the last column in Table I. Green shading corresponds to the stable cw-state. Yellow shading  corresponds to the $|\mu|$-specific instability tongues, i.e., Arnold tongues, intermitting with the intervals of stability. Blue shading  corresponds to the simultaneous instabilities  with different $|\mu|$, and with the cw stability being fully lost.  Dashed red lines show the higher-order thresholds for $3$ and $4$ sideband pairs to be unstable simultaneously.
	}
	\label{f6}
\end{figure*}

Fig.~\ref{f6} shows plots of $\vg_\mu^{(i)}(\vD_\mu(\al_\mu))$  
vs $\al_\mu$, see Eq.~\bref{dc}. 
It also shows the horizontal line corresponding to the minima of the tongues, see Eqs.~\bref{mn}, i.e., to
\be
\vg=\vg_{\mu F}^{(2)}(\vD_\mu=\kappa)=\frac{\kappa}{2}.
\lab{e34aa}
\ee
Fig. \ref{f6} applies only to those ranges of the detunings where the FWM (blue dots) and complexity (red dots) thresholds exist, see Fig.~\ref{f3}. 

The small and large $|\al_\mu|$ cases in Fig.~\ref{f6} can be analysed by means of the asymptotic expansions, leading to the
explicit expressions for $\cW_{\mu C}$ and $\cW_{\mu F}$.
Eq.~\bref{e18a} and Eqs.~\bref{gg} in the respective limits become:
\bsub
\lab{e34}
\begin{align}
|&\al_\mu|\gg \al_{\wh\mu},~\frac{\vD_\mu}{\kappa}=|\al_\mu|-\frac{\al_\mu}{2}+{\cal O}(\al_\mu^{-1}); \\
&\al_\mu\gg \al_{\wh\mu},~\frac{\vg_{\mu C}^{(1)}}{\kappa}=\frac{\vD_\mu}{\kappa}+{\cal O}(\tfrac{\kappa}{\vD_\mu})=
\frac{1}{2}\al_\mu+{\cal O}(\al_\mu^{-1});\\
-&\al_\mu\gg \al_{\wh\mu},~\frac{\vg_{\mu C}^{(2)}}{\kappa}=\frac{\vD_\mu}{3\kappa}+{\cal O}(\tfrac{\kappa}{\vD_\mu})= -\frac{1}{2}\al_\mu+{\cal O}(\al_\mu^{-1}).
\end{align}
\esub
and
\bsub
\lab{e35}
\begin{align}
|\al_\mu|\ll \al_{\wh\mu},~ &\frac{\vD_\mu}{\kappa}=1-\frac{\al_\mu}{2}+\frac{\al_\mu^2}{2}+{\cal O}(\al_\mu^{4});
\\
&\frac{\vg_{\mu C}^{(2)}}{\kappa}=\frac{1}{2}+\frac{\al_\mu^2}{4}+{\cal O}(\al_\mu^4).
\lab{e35b}
\end{align}
\esub
In the limit $\al_\mu\to 0$, $\vg_{\mu C}^{(2)}$ in Eq.~\bref{e35b} naturally 
tends to $\vg_{\mu F}^{(2)}$ in Eq.~\bref{e34aa}.
The same calculations can be repeated with $\al_\mu^{(N)}$, Eq. \bref{npair},
to find thresholds corresponding to the simultaneous instabilities of the $N$ pairs of sidebands. If the approximation discussed after  Eq. \bref{npair} is applied, then the respective $\vg_{\mu C}^{(i)}$ are still solely parametrised by $\al_\mu$,
which is used in Fig.~\ref{f6} to plot $N=3$, and $N=4$ thresholds. 

Fig.~\ref{f6} is, in essence, a summary of our results. It shows that the values of the sideband coupling constant, $\vg$, and the residual finesse, $\al_\mu$, specify three possible regimes of the
resonator operation along the tails of the cw-resonance in Fig.~\ref{f1}: (i)~green area: stable cw-state; (ii)~red area: $|\mu|$-specific instability tongues, i.e., Arnold tongues, intermitting with the intervals of stability; (iii)~blue area: simultaneous instabilities of the tongues with different $|\mu|$, and with the cw stability being fully lost. 

\subsection{Pump power}
Practical approximations for the laser power along 
the threshold of complexity can also  be worked out using that the tongues are
largely located along the tails of the cw-state, where  
$|\delta_0|/\vg\gg 1$, see Fig.~\ref{f3}, and hence, Eq.~\bref{cw} is reduced to
\be
\frac{\cW}{\cW_*}\approx \frac{\vg}{\kappa}~\frac{4\delta_0^2}{\kappa^2}.
\lab{e40}
\ee
Minima of the laser power  are then $\p_{\delta_0}[\cW/\cW_*]\approx 4\delta_0^2[\p_{\delta_0}\vg]\big(1+{\cal O}(\vg/\delta_0)\big)/\kappa^3$, and are closely matched by the minima of $\vg_\mu^{(2)}$, see Eqs.~\bref{mn}.

For $|\mu|\gg 1$, which is also consistent with 
$|\delta_0|/\vg\gg 1$,
\be
\delta_0^{(\mu C)}\approx\delta_0^{(\mu F)}\approx-\tfrac{1}{2}\mu^2D_2,
\lab{e42}
\ee
see Eqs. \bref{df}, \bref{dc}.
Using Eqs. \bref{e34aa} we then find 
\be
\frac{\cW_{\mu F}}{\cW_*}\approx 
\frac{2\delta_0^2}{\kappa^2}
\approx 
\frac{\cF_d^2\mu^4}{2}
\lab{e43}
\ee
for the FWM threshold, and, using Eqs. \bref{e34}, 
\be
\frac{\cW_{\mu C}}{\cW_*}\approx
\frac{2\delta_0^2}{\kappa^2}
\frac{\sq{2|\delta_0D_2|}}{\kappa}
\approx
\frac{ |\cF_d|^3|\mu|^5}{2}
\lab{e44}
\ee
for the threshold of complexity.

Thus,  $|\al_\mu|\approx |\mu\cF_d|$ is the factor making the difference between the FWM and complexity thresholds in the large detuning limit,
\be
\frac{\cW_{\mu C}}{\cW_{\mu F}}\approx|\al_\mu|, ~\frac{|\delta_0|}{\vg}\gg 1.
\ee
The above ratio also  measures the 
relative contrast of the Arnold tongues. 

Let us recall, that $\cW_*\sim\kappa^2$, see Eq.~\bref{ps}. Thus in the dispersive limit, $|\delta_0|\gg\vg$, the FWM threshold power for large $|\mu|$ scales with $D_2^2$, and is $\kappa$ independent in the leading order.
The complexity-threshold power scales inversely with $\kappa$ and is also more sensitive to $D_2$. The $\cW_{\mu C}\sim 1/\kappa$ scaling may look counter-intuitive at first glance, but is in fact natural, since the complexity means the simultaneous instability of different $|\mu|$, which is harder to achieve if the linewidth is narrow, i.e., $|\cF_d|=|D_2|/\kappa$ is large.
In the small $|\delta_0|/\kappa$ limit, the threshold power scales with $\kappa^2$ as is given by $\cW_*$, see Eq.~\bref{con}.

\section{Topical connections and further directions}
The threshold of complexity also has its counterpart in the theory of nonlinear lattices in the condensed matter and optical contexts. Namely, this is the 
Peierls-Nabarro (PN) potential barrier, which constitutes the energy difference between the localised nonlinear lattice excitations centred on one lattice site and the two neighbouring sites~\cite{pn1,pn2}. The latter requires higher energy. Here, we have dealt with the frequency-space 
version of the PN-barrier. Indeed, the threshold of complexity is the two-mode-number, $\pm\mu$, $\pm(\mu+ 1)$, excitation threshold requiring higher energy than the one-mode-number, $\pm\mu$, threshold.

Ref.~\cite{arnold} has made a particular 
emphasis on the synchronisation physics, that has allowed to associate the instability domains in our system with the classic Arnold tongues and to make several cross-disciplinary links \cite{bio}.
While, our work is focused on the parameter range outside the soliton regime, the Arnold-tongue connected soliton physics achieved via synchronisation of the fibre-coupled pair of resonators \cite{gaeta} and via the soliton-breather period synchronising to the repetition rate \cite{papp} has also attracted recent attention.

Arnold tongues in a proximity of the zero dispersion and in the presence of the higher-order dispersion terms, e.g., in connections to 
recent experimental results in Ref.~\cite{f5}, and in the few photon regimes~\cite{quant1,quant2}  emerge as the interesting problems to look at.
Boundaries of the tongues are the bifurcation lines, where new solutions are emerging. These solutions are the Turing roll patterns.
While Refs.~\cite{llprl,csf,tp3,optica}, see also  references therein and thereafter, cover the prior work on Turing rolls in Kerr resonators with continuous or quasi-continuous residual spectra, our Ref. \cite{arnold} has found that  inside the Arnold tongues these structures demonstrate frequency domain symmetry breaking,  and transition to the multimode chaos on approach to the  threshold of complexity. Further understanding of these and search for new properties of the Turing rolls in the high-$\cF_d$ case call for a dedicated near future investigation.

\section{Summary}
We have presented a comprehensive theory of the interplay between the finesse dispersion, $\cF_d=D_2/\kappa$, 
and the frequency conversion thresholds in the Kerr microresonators when
the pump frequency is scanned along the dispersive tails of the nonlinear resonance.
The linear stability analysis of the dispersive tails of the cw-state predicts sharp variations in the threshold power in the resonators with the ultrahigh finesse, $\cF=D_1/\kappa$,
and large $\cF_d$. Large areas of the cw-stability separated by the instability tongues and located above the minimal intra-resonator powers required for FWM come as a surprise, see Figs.~\ref{f3}, \ref{f6}.

Our theory  reveals the key role played by the residual finesse parameter, $\al_\mu\approx\mu\cF_d$, 
in determining the difference between the power thresholds 
for the simultaneous excitation of the two pairs of modes, $\pm\mu$, 
$\pm(\mu+ 1)$, (threshold of complexity) and for one pair, $\pm\mu$ (FWM threshold).
We have determined the critical value of $\cF_d$, i.e., $|\cF_d|\ge 2/\sqrt{3}$, 
for the FWM-free bistability and the pronounced tongue structure to start at the near-zero detunings. 
If $|\cF_d|<2/\sqrt{3}$, then the tongues become pronounced for the detunings  exceeding the critical value, $|\delta_0|/\kappa> 1/6|\cF_d|+\dots$.

Overall, our work underpins  development 
of the microresonator based frequency conversion sources by predicting optimal  
pump laser frequencies and 
powers for achieving mode-number specific conversion in the high-finesse devices.

\section*{Acknowledgement}
We acknowledge discussions with Z. Fan.

\section*{Funding}
EU Horizon 2020 Framework Programme (812818, MICROCOMB).

\section*{Appendix: Parameter estimates}
Estimates for the crystalline, e.g., CaF$_2$ \cite{hq3,f2}, resonators
are $S=100\mu$m$^2$, $n_0=1.43$, $n_2=1.71\times$10$^{-20}$m$^2$/W, $\om_0=200$THz, and hence $\gamma\simeq 10$kHz/W. $2$mm radius corresponds to $D_1/2\pi=16$GHz. $D_2$ can be either positive or negative and order of few kHz, we take $D_2=1$kHz.
The linewidth $\kappa/2\pi=200$kHz corresponds to the quality factor $Q=10^9$ and finesse $\cF\simeq 10^5$. So that $W_*=\pi\kappa^2/\eta D_1\gamma\simeq 1$mW, $\eta=0.5$, $|\cF_d|=0.005$. 
Taking an ultrahigh-Q resonator with 
$\kappa/2\pi=2$kHz, $Q=10^{11}$  gives 
$\cW_*\simeq 0.1\mu$W, and $|\cF_d|=0.5$. 

Estimates for the integrated Si$_3$N$_4$ resonators \cite{hq1,hq2,rev1,revv2}
are $S=0.5\mu $m$^2$, $n_0=2$, $n_2=2.4\times 10^{-19}$m$^2$/W, $\om_0=200$THz, and hence $\gamma\simeq 20$MHz/W. $23\mu$m radius corresponds to $D_1/2\pi=1$THz. $D_2$ can be either positive or negative, we take $D_2=50$MHz 
for the width $\simeq 1.5\mu$m and height $\simeq 0.8\mu$m.
The linewidth $\kappa/2\pi=100$MHz corresponds to the quality factor $Q=2\times 10^6$ and finesse $\cF\simeq  10^4$. So that $W_*\simeq \pi\kappa^2/\eta D_1\gamma\simeq 4$mW, $\eta=0.5$, $|\cF_d|=0.5$. Taking an ultrahigh-Q resonator
with $\kappa/2\pi=10$MHz, $Q=2\times 10^7$  gives 
$\cW_*\simeq 40\mu$W, and $|\cF_d|=5$.
Concise summary of the scaling powers is given in Section II.C, Table I.

\end{document}